\documentclass[aps,pre,superscriptaddress,reprint]{revtex4-2}

\usepackage{graphicx}
\usepackage{amsmath,amssymb,amsfonts}
\usepackage{bm}
\usepackage{xcolor}
\usepackage{hyperref}
\usepackage{booktabs}

\hypersetup{
    colorlinks=true,
    linkcolor=blue,
    urlcolor=magenta,
    citecolor=red,
    pdftitle={
    Active particles trace Lorenz-like butterfly attractors
    in steady multivortex flows
    }
}

\begin{document}

% ============================================================
% TITLE
% ============================================================

%\title{
%Active particles trace Lorenz-like attractors
%in steady multivortex flows
%}

\title{Chaos Made Visible: Active particles trace Lorenz-like attractors in physical space}

% ============================================================
% AUTHORS
% ============================================================
\author{Rahil N. Valani}\email{rahil.valani@physics.ox.ac.uk}
\affiliation{
Rudolf Peierls Centre for Theoretical Physics,
Parks Road,
University of Oxford,
Oxford OX1 3PU,
United Kingdom
}

\author{Ekashman Tiwari}
\affiliation{
Department of Aerospace Engineering,
Indian Institute of Space Science and Technology,
Thiruvananthapuram 695547, Kerala, India
}

\author{Aritra Roy}
\affiliation{
Department of Mechanical Engineering,
Indian Institute of Technology Bombay,
Powai, Mumbai 400076, India
}

\date{\today}

% ============================================================
% ABSTRACT
% ============================================================

\begin{abstract}

Chaotic attractors are typically visualized in an abstract phase space. Here we show that an orientable active particle moving through a steady, two-dimensional, multivortex flow can trace a Lorenz-like chaotic attractor in its real space trajectory. Despite the flow being strictly time independent, coupling between particle position and orientation creates a three-dimensional nonlinear dynamical system. For suitable particle shape and swimming speed, the trajectory organizes into a striking two-lobed geometry reminiscent of the classical Lorenz attractor. A structured first return map, positive largest Lyapunov exponent, and a broadband power spectrum establish that the underlying motion is chaotic. Varying the swimming speed reveals a rich sequence of dynamical regimes as self-propulsion breaks streamline confinement and reorganizes the particle motion. Our results demonstrate how an internal orientational degree of freedom can make chaotic dynamics directly visible in physical space, allowing a single active particle to trace a chaotic attractor as it moves through an otherwise steady flow.

\end{abstract}

\maketitle

% ============================================================
% I. INTRODUCTION
% ============================================================

\section{Introduction}
\label{sec:introduction}

In 1963, Edward Lorenz showed that a simple deterministic model of atmospheric convection could produce bounded, aperiodic motion with extreme sensitivity to initial conditions \cite{Lorenz1963}. Its strange attractor, with the iconic two-lobed ``butterfly'' geometry, became one of the defining images of deterministic chaos. A trajectory winds repeatedly around one lobe before switching irregularly to the other, all within a system of only three nonlinear ordinary differential equations. Yet, the Lorenz butterfly lives in an abstract phase space rather than in physical space. This naturally raises the question of whether similarly recognizable chaotic attractor geometries can be traced directly by the motion of a particle in real space.

Active particles provide a natural setting in which nonlinear dynamics can become directly coupled to physical motion. These systems encompass swimming microorganisms, motile cells, and synthetic microswimmers that convert energy from their surroundings or internal processes into
directed motion \cite{elgeti2015physics,bechinger2016active,KochSubramanian2011,Brumley2015}.Unlike a
passive tracer, which is transported entirely by the surrounding fluid, an active particle immersed in a fluid medium also moves relative to the fluid along its direction of motion. For an anisotropic particle, this orientation is continuously modified by the local velocity gradients through Jeffery dynamics \cite{jeffery1922motion}. Position and orientation are therefore dynamically coupled. The particle position determines the flow gradients it experiences, while its orientation determines how self-propulsion carries it through the flow.

This feedback can generate rich nonlinear motion even when the background flow itself is laminar and time independent. Active particles in shear and Poiseuille flows can exhibit swinging, tumbling, trapping, and other exotic trajectories \cite{zottl2012nonlinear,Valani2024,Rusconi2014,Harding2025,Borgnino2022,Zottl2013}. Vortical flows are particularly interesting because self-propulsion and flow-induced reorientation compete with circulation around the underlying vortices. Deterministic trapping and escape have been studied for microswimmers in elementary vortical flows
\cite{Durham2011,Sokolov2016,tanasijevic2022microswimmers,
Chajwa2026}, while self-propulsion can modify transport barriers and produce preferential transport and aggregation in laminar flows \cite{torney2007transport,Ghosh2021,GhoshDebnath2021}. Related nonlinear phenomena also arise for passive inertial particles in vortical flows, where particle dynamics can produce trapping, bifurcations, limit cycles, and chaotic attractors \cite{Kapoor2024}.

A crucial consequence of this coupling is that motion in a two-dimensional steady flow does not imply a two-dimensional particle dynamical system. An orientable active particle is described by two spatial coordinates together with its orientation, so that its complete state evolves as a three-dimensional autonomous system. This additional degree of freedom makes continuous-time deterministic chaos possible even though the underlying velocity field is steady. Chaotic active-particle dynamics have been reported
in steady channel and imposed flows
\cite{Valani2024,Richardson2018,Truneh2026-bu},
and recent experiments on swimming microbes in a
time-independent vortex-chain flow have demonstrated
chaotic motion, dynamical trapping, and anomalous transport
\cite{Le2025}. Temporal complexity of the background flow is therefore not required for chaotic motion of active particles.

Here we uncover a striking geometrical manifestation of this dynamics. An active particle moving through a steady multivortex flow can trace a Lorenz-like butterfly attractor directly in the two-dimensional physical plane. For suitable particle shape and swimming speed, the trajectory organizes into two spatially separated lobes. The particle circulates
repeatedly within each lobe and switches irregularly between them, producing a real-space pattern reminiscent of the classical Lorenz butterfly. The resemblance is geometrical rather than an equivalence between the two dynamical systems. The remarkable feature is that the two-lobed organization of the dynamics becomes directly visible in its projection onto physical space.

The remainder of the paper is organized as follows. Section~\ref{sec:model} introduces the steady Magnetohydrodynamics (MHD) multivortex flow and the active-particle model. Section~\ref{sec:butterfly} presents the physical-space attractors and the emergence of the Lorenz-like two-lobed geometry. Section~\ref{sec:chaos} characterizes the associated chaotic dynamics using return maps, Lyapunov exponents, and spectral analysis. Section~\ref{sec:activity} examines how the particle dynamics reorganize as the swimming speed is varied. Finally, Sec.~\ref{sec:discussion} summarizes the main findings and discusses their broader implications and
possible extensions.

% ============================================================
% II. MODEL
% ============================================================

\section{Model}
\label{sec:model}

We consider a two-dimensional electrically conducting fluid confined to a square domain in the $(x,z)$ plane. A time-periodic electromagnetic forcing generates the steady multivortex flow in which the active particle moves. The imposed electric field is directed perpendicular to the flow plane and is written in dimensional form as
\begin{equation*}
    \mathbf{E}(x,z,t)
    =
    \Re\!
    \left\{
    E_y(x,z)e^{i\omega t}
    \right\}\mathbf{e}_y ,
  %  \label{eq:electric_field}
\end{equation*}
where $\omega$ is the forcing frequency. Its spatial
amplitude is prescribed by two oppositely signed Gaussian
regions,
\begin{equation*}
\begin{split}
    E_y(x,z)
    =
    E_0
    \Bigg[
    &
    \exp\left(
    -\frac{(x-x_1)^2+(z-z_1)^2}{2w^2}
    \right)
    \\
    &-
    \exp\left(
    -\frac{(x-x_2)^2+(z-z_2)^2}{2w^2}
    \right)
    \Bigg]
    e^{-i\beta_E z},
\end{split}
%\label{eq:forcing_profile}
\end{equation*}
where $E_0$ sets the field amplitude, $w$ the Gaussian
width, $\beta_E$ the spatial wavenumber, and
$(x_1,z_1)$ and $(x_2,z_2)$ their centers, located along a
diagonal of the square domain. The associated electric current and
induced magnetic field produce a time-averaged Lorentz
force $\langle\mathbf{F}_L\rangle$~\cite{vynnycky2022fast} whose spatial curl drives
the multivortex pattern shown in
Fig.~\ref{fig:butterfly}(a). Details of the electromagnetic
formulation are given in Appendix~\ref{app:mhd}.

At low Reynolds number, fluid inertia is negligible and the
dimensional velocity field
\begin{equation*}
    \mathbf{u}(x,z)
    =
    \left(u(x,z),w(x,z)\right)
 %   \label{eq:velocity_dim}
\end{equation*}
satisfies the incompressible Stokes equations
%
%\begin{subequations*}
%\label{eq:stokes_dim}
\begin{align*}
    \nabla\cdot\mathbf{u}
    &=0,
    \\
    -\mu\nabla^2\mathbf{u}
    +\nabla p
    &=
    \langle\mathbf{F}_L\rangle ,
\end{align*}
%\end{subequations*}
%
where $\mu$ is the dynamic viscosity and $p$ is the
pressure. The Lorentz force is averaged over the rapid
electromagnetic forcing cycle, and the flow is solved to a
steady state before the particle dynamics are evolved.
Thus, the particle moves through a spatially structured but
strictly time-independent velocity field, and any temporal
complexity in its trajectory arises from the particle
dynamics.

Into this steady multivortex flow we introduce a
self-propelled spheroidal particle with position $(x,z)$
and orientation angle $\theta$. The particle swims with
constant speed $V_0$ relative to the surrounding fluid
along
\begin{equation*}
    \mathbf{p}
    =
    (\cos\theta,\sin\theta).
%    \label{eq:orientation}
\end{equation*}
Its shape is described by the Jeffery parameter
\cite{jeffery1922motion}
\begin{equation*}
    \alpha
    =
    \frac{\beta^2-1}{\beta^2+1},
    \qquad
   % \beta
   % =
   % \frac{l_\parallel}{l_\perp},
%    \label{eq:alpha}
\end{equation*}
where $\beta$ is the aspect ratio. The limits
$\alpha>0$, $\alpha=0$, and $\alpha<0$ correspond to
prolate, spherical, and oblate particles, respectively.

The particle is advected by the local flow while
simultaneously swimming along its orientation, giving
\begin{subequations}
\label{eq:translation_dim}
\begin{align}
    \frac{dx}{dt}
    &=
    u(x,z)+V_0\cos\theta ,
    \label{eq:xdot_dim}
    \\
    \frac{dz}{dt}
    &=
    w(x,z)+V_0\sin\theta .
    \label{eq:zdot_dim}
\end{align}
\end{subequations}
The same flow also reorients the particle through the local
velocity gradients. For the two-dimensional geometry
considered here, Jeffery's equation reduces to
\begin{align}
    \frac{d\theta}{dt}
    &=
    (1+\alpha)\frac{\omega}{2}
    \nonumber\\
    &\quad
    -
    \alpha
    \left[
    2\frac{\partial u}{\partial x}
    \cos\theta\sin\theta
    -
    \frac{\partial u}{\partial z}
    \cos^2\theta
    +
    \frac{\partial w}{\partial x}
    \sin^2\theta
    \right],
    \label{eq:theta_dim}
\end{align}
where
\begin{equation*}
    \omega
    =
    \frac{\partial w}{\partial x}
    -
    \frac{\partial u}{\partial z}
%    \label{eq:vorticity_dim}
\end{equation*}
is the out-of-plane vorticity. All flow quantities are
evaluated at the instantaneous particle position.

Equations~(\ref{eq:translation_dim}) and
(\ref{eq:theta_dim}) establish the feedback that is central to  the dynamics. The particle position determines the local flow gradients that set its orientation, while the orientation determines the direction of self-propulsion and hence where the particle moves next.

\begin{figure*}[t]
\centering
\includegraphics[width=0.85\textwidth]{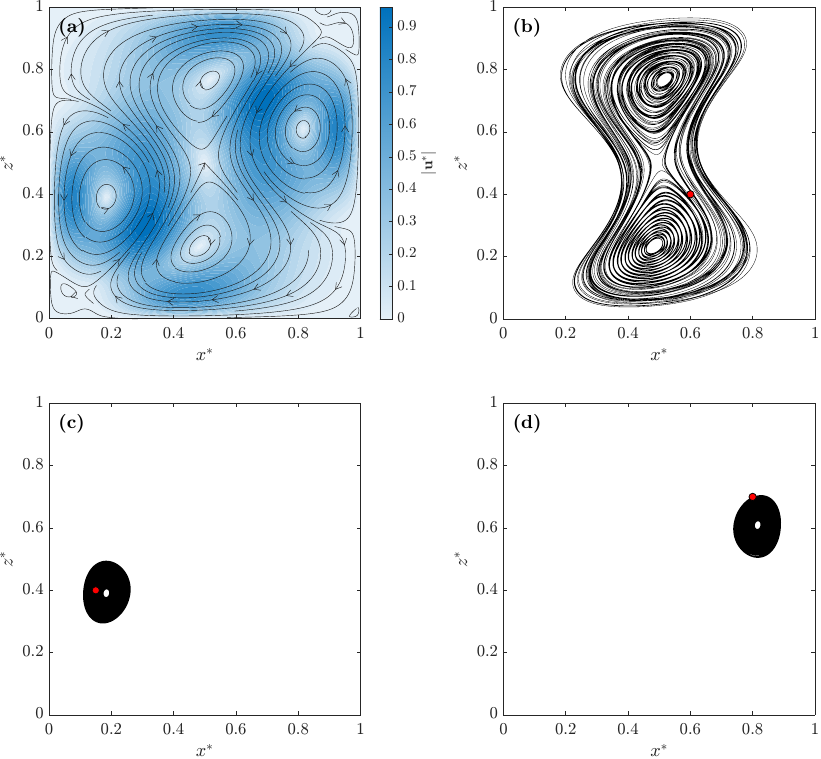}

\caption{
\textbf{Lorenz-like attractor geometry in a steady
two-dimensional multivortex flow.}
(a) Steady MHD background flow, showing the dimensionless
velocity magnitude $|\mathbf{u}^*|$ and streamlines.
(b--d) Physical-space trajectories for identical particle
parameters, $\alpha=-0.8$ and $v_0=0.015$, but different
initial conditions, marked by red circles.
The trajectory in (b) organizes into a two-lobed
Lorenz-like attractor geometry (see also Movie 1).
The initial conditions in (c) and (d) instead approach
distinct localized long-time states, demonstrating the
coexistence of qualitatively different particle dynamics
within the same steady background flow.
}
\label{fig:butterfly}
\end{figure*}

We nondimensionalize lengths by the characteristic length
$L_c$, velocities by the characteristic flow speed $U_c$,
and time by the advective scale $L_c/U_c$. Thus,
\begin{equation}
    x^*=\frac{x}{L_c},
    \quad
    z^*=\frac{z}{L_c},
    \quad
    t^*=\frac{U_c}{L_c}t,
    \quad
    \mathbf{u}^*=\frac{\mathbf{u}}{U_c}.
    \label{eq:nondim}
\end{equation}
The corresponding dimensionless swimming speed is
\begin{equation}
    v_0=\frac{V_0}{U_c},
    \label{eq:v0}
\end{equation}
which measures the strength of self-propulsion relative to
advection by the background flow.

In these variables, the particle dynamics become
\begin{subequations}
\label{eq:particle}
\begin{align}
    \frac{dx^*}{dt^*}
    &=
    u^*(x^*,z^*)
    +v_0\cos\theta ,
    \label{eq:xdot}
    \\
    \frac{dz^*}{dt^*}
    &=
    w^*(x^*,z^*)
    +v_0\sin\theta ,
    \label{eq:zdot}
    \\
    \frac{d\theta}{dt^*}
    &=
    (1+\alpha)\frac{\omega^*}{2}
    \nonumber\\
    &\quad
    -
    \alpha
    \left[
    2\frac{\partial u^*}{\partial x^*}
    \cos\theta\sin\theta
    -
    \frac{\partial u^*}{\partial z^*}
    \cos^2\theta
    +
    \frac{\partial w^*}{\partial x^*}
    \sin^2\theta
    \right],
    \label{eq:thetadot}
\end{align}
\end{subequations}
where
\begin{equation*}
    \omega^*
    =
    \frac{\partial w^*}{\partial x^*}
    -
    \frac{\partial u^*}{\partial z^*}
  %  \label{eq:vorticity}
\end{equation*}
is the dimensionless vorticity. These
equations, together with the prescribed steady velocity
field $\mathbf{u}^*(x^*,z^*)$, form the dynamical system
studied below.

Although the background flow is two-dimensional and steady,
the particle has the three-dimensional state
\begin{equation*}
    \mathbf{X}
    =
    (x^*,z^*,\theta)^T,
  %  \label{eq:state}
\end{equation*}
so that Eqs.~(\ref{eq:particle}) define the autonomous
system
\begin{equation*}
    \dot{\mathbf{X}}
    =
    \mathbf{F}(\mathbf{X}),
%    \label{eq:3Dsystem}
\end{equation*}
where the dot denotes differentiation with respect to
$t^*$. %The orientational degree of freedom therefore
%elevates particle motion in a steady two-dimensional flow
%to a three-dimensional dynamical system capable of
%supporting complex dynamics.

Importantly, incompressibility of the fluid does not imply
volume preservation in this enlarged particle state space.
Using $\nabla^*\cdot\mathbf{u}^*=0$, the divergence of
Eqs.~(\ref{eq:particle}) is
\begin{equation*}
\begin{split}
\nabla_{\mathbf{X}}\cdot\dot{\mathbf{X}}
=
-\alpha
\Bigg[
&
2\frac{\partial u^*}{\partial x^*}
\cos(2\theta)
+
\left(
\frac{\partial u^*}{\partial z^*}
+
\frac{\partial w^*}{\partial x^*}
\right)
\sin(2\theta)
\Bigg].
\end{split}
%\label{eq:phase_divergence}
\end{equation*}
For nonspherical particles $(\alpha\neq0)$, this quantity
is generally nonzero, allowing local contraction and
expansion of volumes in $(x^*,z^*,\theta)$ space. The
particle dynamics can therefore support attracting sets
despite the incompressibility of the underlying flow. In
the spherical limit $\alpha=0$, the smooth dynamics are
instead volume preserving.

The steady flow is computed on a uniform Cartesian grid,
with the velocity and its gradients evaluated at the
instantaneous particle position by bilinear interpolation.
Particle trajectories are integrated using a fourth-order
Runge--Kutta scheme with $\Delta t^*=10^{-3}$, with elastic
reflection at the domain boundaries. Unless otherwise
stated, all trajectories are initialized at
$(x_0^*,z_0^*,\theta_0)=(0.6,0.4,0)$.

% ============================================================
% III. PHYSICAL-SPACE BUTTERFLY ATTRACTOR
% ============================================================

\section{Physical-space attractors}
\label{sec:butterfly}

Figure~\ref{fig:butterfly} captures the central result of this work. The steady MHD flow in
Fig.~\ref{fig:butterfly}(a) forms spatially structured vortices, yet the active particle trajectory can organize into something remarkably different. For $\alpha=-0.8$ and $v_0=0.015$, the particle traces the striking two-lobed structure shown in Fig.~\ref{fig:butterfly}(b), reminiscent of the classical Lorenz  (see Movie 1).

The trajectory is organized around two well-defined spatial lobes in the upper and lower parts of the domain. Within each lobe, the particle executes repeated rotations, but no two excursions are quite the same. The size and shape of successive loops vary irregularly, with the trajectory continually revisiting regions without settling onto a closed orbit. Over time, these nested and nonrepeating excursions build up the characteristic Lorenz-attractor-shaped geometry.

What makes this structure unusual is that it is visible directly in the physical $(x^*,z^*)$ plane. The classical Lorenz butterfly is an attractor in an abstract three-dimensional phase space \citep{Lorenz1963}. Here, the full particle dynamics likewise evolve in the three-dimensional state space $(x^*,z^*,\theta)$, yet their projection onto physical space retains a remarkably clear two-lobed organization. We therefore use the term ``Lorenz-like'' to describe this geometrical resemblance, without implying equivalence to the Lorenz equations. Moreover, the visual complexity alone is not sufficient to establish chaos, since irregular-looking projections can also arise from regular higher-dimensional dynamics.

The other trajectories in Fig.~\ref{fig:butterfly} reveal
an equally important feature. The flow and particle
parameters are identical in panels (b--d), and only the
initial state is changed. Nevertheless, the long-time
motion is markedly different. While panel (b) produces the
extended two-lobed attractor, the trajectories in panels
(c) and (d) approach localized states in different
regions of the domain. The same steady flow therefore
supports multiple coexisting attractors with very different
physical-space geometries.

% This sensitivity to the initial state also makes clear that
% the observed structures are not simply imprints of the
% underlying streamlines. Instead, they emerge from the
% nonlinear interplay of advection, self-propulsion, and
% flow-induced reorientation. A single time-independent flow
% can therefore organize active-particle motion into
% qualitatively distinct long-time states, including the
% extended Lorenz-like structure.

\begin{figure*}[t]
\centering
\includegraphics[width=\textwidth]{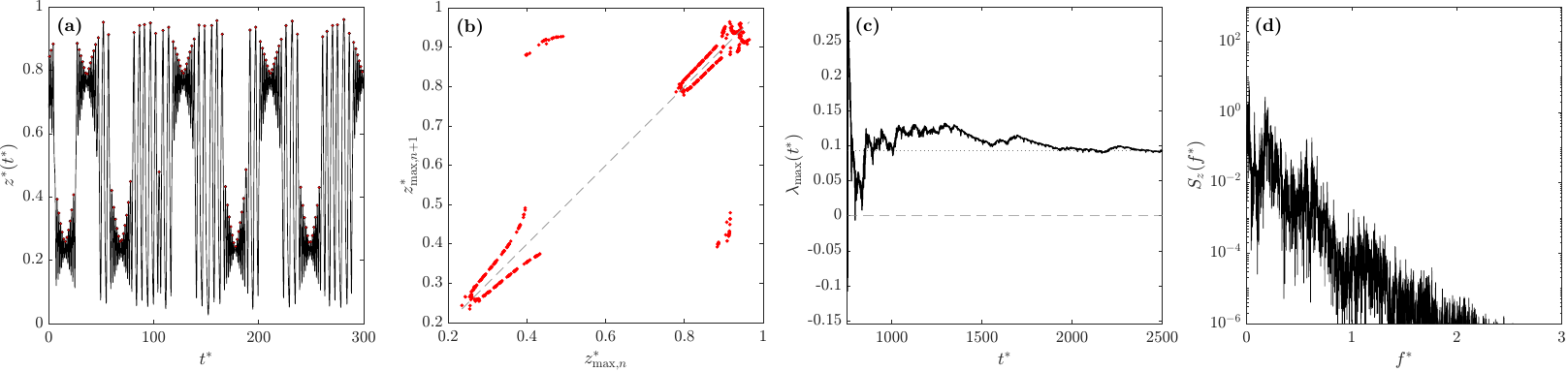}

\caption{
\textbf{Characterization of the chaotic dynamics.}
(a) Representative time series $z^*(t^*)$, with red circles
marking successive local maxima.
(b) Peak-to-peak return map
$z_{\max,n+1}^*$ versus $z_{\max,n}^*$; the dashed line
denotes $z_{\max,n+1}^*=z_{\max,n}^*$.
(c) Running estimate of the largest Lyapunov exponent,
which converges to a positive value. The dashed horizontal
line marks $\lambda_{\max}=0$, and the dotted line indicates
the asymptotic estimate.
(d) Power spectral density $S_z(f^*)$ of $z^*(t^*)$.
The structured return map, positive largest Lyapunov
exponent, and broadband spectrum together identify the dynamics as chaotic.
}
\label{fig:chaos}
\end{figure*}

% ============================================================
% IV. CHAOTIC BUTTERFLY DYNAMICS
% ============================================================

\section{Characterization of chaotic dynamics}
\label{sec:chaos}

The two-lobed geometry is strongly suggestive of chaotic
motion, but geometry alone is not sufficient to establish
chaos. We therefore further probe the dynamics of the trajectory in
Fig.~\ref{fig:butterfly}(b).

Figure~\ref{fig:chaos}(a) shows a representative segment of
the vertical position time series $z^*(t^*)$. The signal is clearly
oscillatory, yet neither its amplitude nor the sequence of
oscillations repeats regularly. Successive peaks vary in
height, reflecting the nonrepeating excursions that build
the nested loops of the physical-space attractor. To expose
the recurrence underlying these oscillations, we record the
successive maxima $z_{\max,n}^*$ and plot
$z_{\max,n+1}^*$ against $z_{\max,n}^*$.

The resulting first-return map in Fig.~\ref{fig:chaos}(b) reveals a clear combination of structure and irregularity. The points neither collapse onto a finite set, as expected for a simple periodic orbit, nor scatter randomly across the plane. Instead, successive maxima organize into several distinct branches, indicating structured but aperiodic recurrence. Interestingly, this form differs from the characteristic cusp-like return map associated with the classical Lorenz attractor. Thus, while the physical-space trajectory bears a striking geometrical resemblance to the Lorenz butterfly, its recurrence structure is distinct. 

A more direct signature of chaos is provided by the largest Lyapunov exponent, which measures sensitivity to nearby initial conditions. We estimate $\lambda_{\max}$ by evolving a reference trajectory together with a nearby perturbed trajectory in the full $(x^*,z^*,\theta)$ state space. Their separation is measured at fixed time intervals and the perturbation is repeatedly renormalized, with the periodicity of $\theta$ accounted for using the shortest angular separation. After discarding the initial transient, the accumulated separation rates give the running estimate shown in Fig.~\ref{fig:chaos}(c). Its convergence to a positive value demonstrates exponential sensitivity to initial conditions and provides direct evidence of chaotic dynamics.

The same picture emerges in the frequency domain.
Figure~\ref{fig:chaos}(d) shows that the power spectrum of
$z^*(t^*)$ contains substantial broadband content rather
than only the discrete peaks characteristic of periodic
motion. Taken together, the structured return map, positive
largest Lyapunov exponent, and broadband spectrum establish
that the irregular dynamics underlying the Lorenz-like
physical-space attractor are deterministic and chaotic.

\begin{figure*}[t]
\centering
\includegraphics[width=0.95\textwidth]{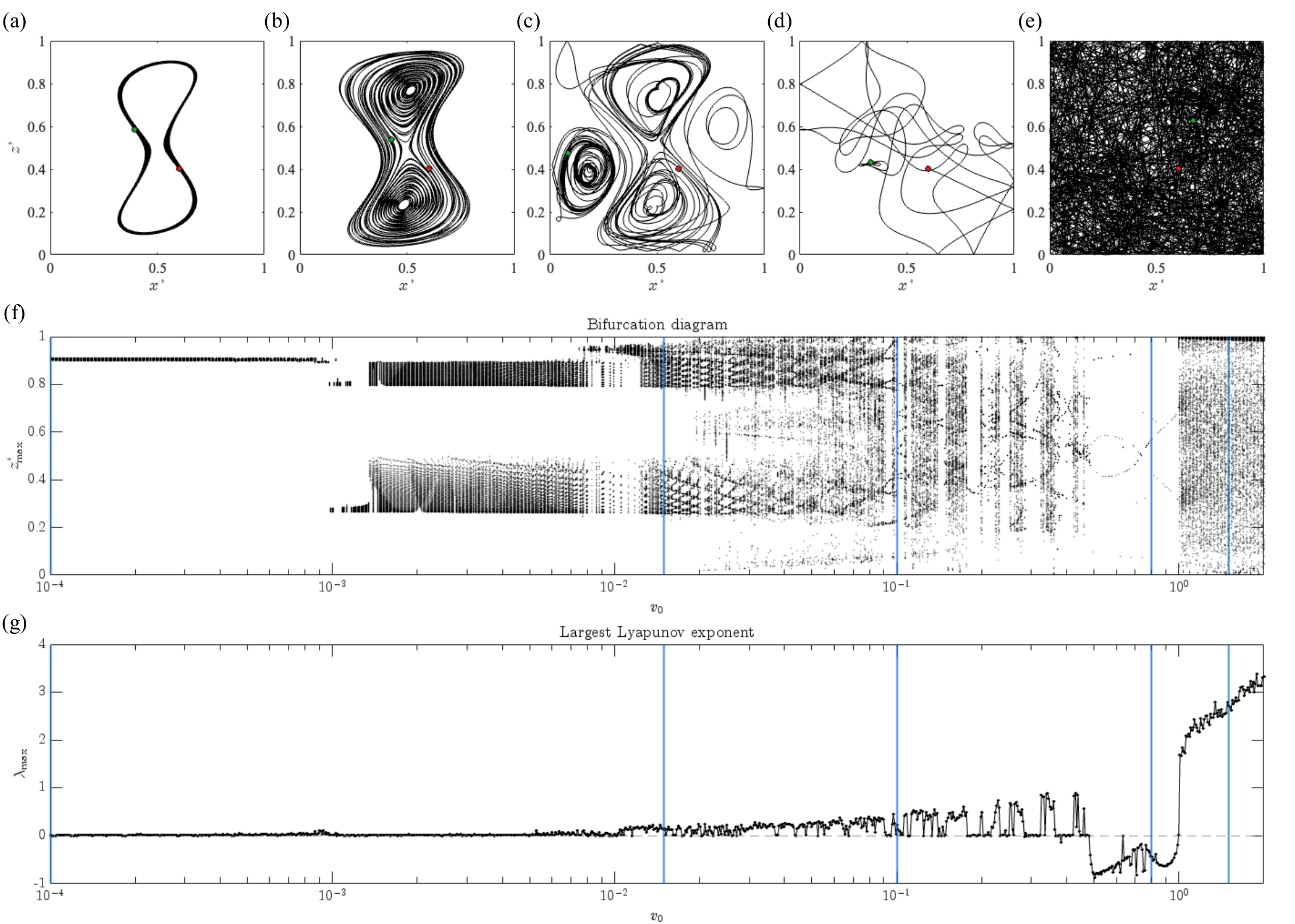}

\caption{
\textbf{Activity-dependent particle dynamics.}
(a--e) Representative physical-space trajectories at fixed
shape parameter $\alpha=-0.8$ and increasing swimming speed,
with (a) $v_0=10^{-4}$,
(b) $v_0=0.015$,
(c) $v_0=0.1$,
(d) $v_0=0.8$, and
(e) $v_0=1.5$.
The red and green circles mark the initial and final particle
positions, respectively.
At weak activity the motion remains strongly organized by
the background flow, while at $v_0=0.015$ the
trajectory develops the two-lobed Lorenz-like geometry.
Increasing activity further produces qualitatively different
trajectories spanning multiple vortical regions and,
eventually, much of the domain.
(f) Bifurcation diagram constructed from the local maxima
$z_{\max}^*$ as a function of $v_0$.
(g) Corresponding largest Lyapunov exponent
$\lambda_{\max}$, with the dashed horizontal line marking
$\lambda_{\max}=0$.
The blue vertical lines in (f) and (g) identify the five
values of $v_0$ represented in panels (a--e).
}
\label{fig:activity}
\end{figure*}

% ============================================================
% V. ACTIVITY REORGANIZES THE PARTICLE DYNAMICS
% ============================================================

\section{Activity reorganizes the particle dynamics}
\label{sec:activity}

The physical-space dynamics result from a competition between advection by the steady background flow and self-propulsion of the particle. To examine how this competition changes the long-time behavior, we fix the particle shape at $\alpha=-0.8$ and vary the swimming speed $v_0$ over more than four orders of magnitude. Figure~\ref{fig:activity} combines representative physical-space trajectories with a bifurcation diagram and the corresponding largest Lyapunov exponent.

We first characterize the global dependence on activity
through the bifurcation diagram in
Fig.~\ref{fig:activity}(f). For each value of $v_0$, the
particle is evolved beyond an initial transient and the
successive local maxima $z_{\max}^*$ of its vertical
position are recorded. A periodic trajectory therefore
produces a finite number of discrete branches, whereas
irregular dynamics generate a broader set of sampled
maxima. This representation provides a compact view of how
the long-time particle dynamics reorganize as
self-propulsion is increased. The corresponding largest
Lyapunov exponent in Fig.~\ref{fig:activity}(g) provides an
independent measure of this transition, with
$\lambda_{\max}>0$ identifying parameter ranges with
sensitive dependence on initial conditions.

In the passive limit $v_0=0$, the translational dynamics
reduce to
\begin{equation*}
    \dot{x}^*=u^*,
    \qquad
    \dot{z}^*=w^* .
%    \label{eq:passive_limit}
\end{equation*}
For a steady two-dimensional incompressible flow, these
equations form a one-degree-of-freedom Hamiltonian system,
with the streamfunction $\psi^*$ acting as the Hamiltonian.
Thus $\psi^*$ is conserved and passive trajectories remain
confined to streamlines. At small but finite $v_0$,
self-propulsion weakly perturbs this streamline confinement. Accordingly, at $v_0=10^{-4}$
[Fig.~\ref{fig:activity}(a)] the trajectory remains strongly
organized by the background flow, the bifurcation diagram
contains well-defined branches, and $\lambda_{\max}$ remains
close to zero.

As $v_0$ increases, self-propulsion allows the particle to
depart progressively from passive streamline confinement.
The bifurcation diagram first develops additional branches
and subsequently broadens, indicating increasingly complex
long-time dynamics. At $v_0=0.015$,
Fig.~\ref{fig:activity}(b), this competition produces the
organized two-lobed physical-space attractor discussed in
Secs.~\ref{sec:butterfly} and \ref{sec:chaos}. At this
activity, the steady vortex structure continues to organize
the trajectory spatially, while self-propulsion and
flow-induced reorientation generate nonperiodic motion.
The positive largest Lyapunov exponent confirms that this
regime is chaotic.

Increasing the swimming speed further produces a much
broader range of physical-space dynamics. At
$v_0=0.1$, Fig.~\ref{fig:activity}(c), the particle is
no longer confined to the two-lobed structure and instead
explores several vortical regions of the background flow. This
change is reflected in the substantial broadening and
branching of the $z_{\max}^*$ values in
Fig.~\ref{fig:activity}(f). Over much of this intermediate
range, $\lambda_{\max}$ remains positive, showing that
chaotic dynamics are not restricted to the particular
parameter value at which the Lorenz-like geometry is most
clearly visible.

Importantly, the dependence on activity is not monotonic.
The bifurcation diagram contains windows in which the
chaotic dynamics give way to more ordered motion and the
largest Lyapunov exponent approaches zero or becomes
negative. At $v_0=0.8$, for example,
$\lambda_{\max}<0$ and the trajectory in
Fig.~\ref{fig:activity}(d) approaches a stable fixed point,
with nearby trajectories contracting toward the same
asymptotic state. Increasing self-propulsion therefore does
not simply drive the system from regular to progressively
more chaotic motion. Instead, distinct dynamical states,
including periodic, chaotic, and stable fixed-point
behavior, emerge as the balance between swimming and
advection is varied.

At still stronger activity the dynamics reorganize once
again. For $v_0=1.5$, Fig.~\ref{fig:activity}(e), the
particle explores a substantial fraction of the physical
domain. The bifurcation diagram becomes densely populated
and the largest Lyapunov exponent is strongly positive,
indicating a return to strongly chaotic dynamics in which
self-propulsion plays a much greater role relative to the
underlying vortex structure.

Figure~\ref{fig:activity} therefore reveals a progression
from nearly Hamiltonian, streamline-organized motion at
weak activity to a sequence of branched, chaotic, and
contracting dynamical regimes as self-propulsion is
increased. The Lorenz-like physical-space attractor appears
within an intermediate range where the two competing
effects are comparable. Self-propulsion is sufficiently
strong to break passive streamline confinement, while the
steady multivortex flow remains sufficiently influential to
organize the chaotic trajectory into a coherent two-lobed
geometry. At larger $v_0$, this balance is lost and other
forms of particle motion emerge.

% ============================================================
% VI. DISCUSSION AND CONCLUSION
% ============================================================

\section{Discussion and conclusion}
\label{sec:discussion}

We have shown that an orientable active particle moving
through a steady two-dimensional multivortex flow can trace
a Lorenz-like strange attractor directly in physical space.
Its two-lobed trajectory provides a striking real-space
signature of the underlying chaotic dynamics, effectively
allowing the particle to draw its own chaotic attractor.

More broadly, this behavior illustrates how an internal
degree of freedom can fundamentally enrich motion through a
temporally steady environment. Here, orientation couples
the local velocity gradients to self-propelled translation,
creating dynamics that extend beyond those of a passive
tracer. Such coupling provides a general mechanism by which
complex attractors can emerge in otherwise steady
environments and acquire recognizable geometries in the
trajectories they produce.

The mechanism is not specific to MHD forcing, since the
particle responds to the local velocity and velocity
gradients rather than to how the flow is generated. This
raises the broader question of how physical-space
attractors depend on flow topology, particle shape,
activity, and initial conditions. Such dependence may also
offer a route to controlling active-particle transport,
with particle or flow parameters used to select between
localized, extended, and chaotic trajectories without
requiring an unsteady background flow.

An important future direction is to establish how broadly
physical-space attractors arise across different particle
properties, flow geometries, and levels of environmental
noise. Extending the problem to interacting particles and
three-dimensional flows may reveal still richer forms of
spatially organized chaos. More fundamentally, it will be
important to identify the conditions under which
higher-dimensional nonlinear dynamics retain a simple and
recognizable signature when projected into physical space,
and whether such geometries can be predicted from the
underlying flow structure. Answering this could open a route
to designing steady flows that sculpt particle trajectories
into prescribed dynamical states, turning physical space
itself into a canvas on which complex dynamics become
visible.

% ============================================================
% ACKNOWLEDGMENTS
% ============================================================

 \begin{acknowledgments}
R.V. acknowledges the support of the Leverhulme Trust [Grant No. LIP-2020-014] and of the ERC Advanced Grant ActBio (funded as UKRI Frontier Research Grant EP/Y033981/1).
 \end{acknowledgments}

% ============================================================
% REFERENCES
% ============================================================

% ============================================================
% APPENDIX
% ============================================================

\appendix

\section{Electromagnetic forcing and steady background flow}
\label{app:mhd}

Here we summarize the electromagnetic forcing used to
generate the steady multivortex background flow. The fluid is
electrically conducting, with electrical conductivity
$\sigma$ and magnetic permeability $\mu_m$, and occupies a
square domain of characteristic size $L_c$ in the $(x,z)$
plane.

We impose a time-harmonic electric field perpendicular
to the flow plane,
\begin{equation}
    \mathbf{E}(x,z,t)
    =
    \Re\!
    \left\{
    E_y(x,z)e^{i\omega t}
    \right\}\mathbf{e}_y ,
    \label{eq:appendix_E}
\end{equation}
where $\Re\{\cdot\}$ denotes the real part, $\omega$ is the
angular frequency of the electromagnetic forcing, and
$\mathbf{e}_y$ is the unit vector normal to the $(x,z)$
plane. The complex spatial amplitude is prescribed as
\begin{equation}
\begin{split}
    E_y(x,z)
    =
    E_0
    \Bigg[
    &
    \exp\left(
    -\frac{(x-x_1)^2+(z-z_1)^2}{2w^2}
    \right)
    \\
    &-
    \exp\left(
    -\frac{(x-x_2)^2+(z-z_2)^2}{2w^2}
    \right)
    \Bigg]
    e^{-i\beta_E z}.
\end{split}
\label{eq:appendix_Eprofile}
\end{equation}
The two Gaussian forcing regions have equal width $w$ and
opposite sign. Their centers are placed symmetrically about
the center of the square along a line inclined at
$\vartheta=\pi/4$,
%
%\begin{subequations}
\begin{align*}
    x_1
    &=
    x_c-\frac{d}{2}\cos\vartheta,
    &
    z_1
    &=
    z_c-\frac{d}{2}\sin\vartheta,
    \\
    x_2
    &=
    x_c+\frac{d}{2}\cos\vartheta,
    &
    z_2
    &=
    z_c+\frac{d}{2}\sin\vartheta ,
\end{align*}
%\end{subequations}
%
where $(x_c,z_c)$ is the domain center and $d$ is the
separation between the forcing centers.

For a harmonic field, we represent the induced magnetic
field~\citep{Griffiths_2017} using an out-of-plane magnetic vector potential
\begin{equation*}
    \mathbf{A}
    =
    A_y(x,z)\mathbf{e}_y .
\end{equation*}
The dimensionless complex amplitude satisfies the
Helmholtz equation
\begin{equation}
    \left(
    \nabla^{*2}
    -
    i\Omega
    \right)A_y^*
    =
    -E_y^*,
    \label{eq:helmholtz_A}
\end{equation}
where
\begin{equation}
    \Omega
    =
    \omega\mu_m\sigma L_c^2
    \label{eq:Omega_em}
\end{equation}
is the dimensionless electromagnetic frequency. The Helmholtz problem is solved with homogeneous Dirichlet
conditions, $A_y^*=0$, on the boundary of the square
domain.

The complex electric-current amplitude is then
\begin{equation}
    J_y^*
    =
    E_y^*
    -
    i\Omega A_y^*,
    \label{eq:current}
\end{equation}
while the in-plane magnetic field follows from
$\mathbf{B}=\nabla\times\mathbf{A}$,
\begin{equation}
    B_x^*
    =
    -\frac{\partial A_y^*}{\partial z^*},
    \qquad
    B_z^*
    =
    \frac{\partial A_y^*}{\partial x^*}.
    \label{eq:magnetic_field}
\end{equation}

Averaging the Lorentz force over one forcing period~\cite{vynnycky2022fast} gives
\begin{equation}
    \left\langle
    \mathbf{F}_L^*
    \right\rangle
    =
    \frac{1}{2}
    \Re
    \left\{
    \mathbf{J}^*
    \times
    \mathbf{B}^{*\,\dagger}
    \right\},
    \label{eq:lorentz_average}
\end{equation}
where $\dagger$ denotes complex conjugation. Since the
current is directed along $y$, its two in-plane components
are
\begin{subequations}
\label{eq:lorentz_components}
\begin{align}
    \left\langle F_{L,x}^*\right\rangle
    &=
    \frac{1}{2}
    \Re
    \left(
    J_y^* B_z^{*\,\dagger}
    \right),
    \\
    \left\langle F_{L,z}^*\right\rangle
    &=
    -\frac{1}{2}
    \Re
    \left(
    J_y^* B_x^{*\,\dagger}
    \right).
\end{align}
\end{subequations}

The curl of this Lorentz force acts as the source of
vorticity in the two-dimensional Stokes flow. Introducing
the streamfunction $\psi^*$ through
\begin{equation}
    u^*
    =
    \frac{\partial\psi^*}{\partial z^*},
    \qquad
    w^*
    =
    -\frac{\partial\psi^*}{\partial x^*},
    \label{eq:streamfunction}
\end{equation}
and defining
\begin{equation*}
    \omega_f^*
    =
    \frac{\partial w^*}{\partial x^*}
    -
    \frac{\partial u^*}{\partial z^*},
\end{equation*}
the steady flow problem may be written as
\begin{subequations}
\label{eq:stream_vorticity}
\begin{align}
    \nabla^{*2}\psi^*
    &=
    -\omega_f^*,
    \\
    \nabla^{*2}\omega_f^*
    &=
    -N
    \left(
    \frac{\partial
    \langle F_{L,z}^*\rangle}{\partial x^*}
    -
    \frac{\partial
    \langle F_{L,x}^*\rangle}{\partial z^*}
    \right).
\end{align}
\end{subequations}
No-slip conditions are imposed at the walls. The resulting
stationary solution forms the multivortex background flow
shown in Fig.~\ref{fig:butterfly}(a).

For the calculations reported here, the dimensionless
forcing parameters are
\begin{equation*}
    w^*=0.15,
    \quad
    \beta_E^*=\pi,
    \quad
    \vartheta=\frac{\pi}{4},
    \quad
    N=1.
\end{equation*}
The forcing centers are symmetrically positioned about
$(x_c^*,z_c^*)=(0.5,0.5)$ with separation
\begin{equation*}
    d^*
    =
    \sqrt{0.5^2+0.5^2}.
\end{equation*}
The dimensional electromagnetic parameters used in the
numerical construction are
\begin{equation*}
    \mu_m=4\pi\times10^{-7},
    \quad
    \sigma=4.486,
    \quad
    L_c=5\times10^{-3}\ {\rm m},
\end{equation*}
together with forcing frequency
\begin{equation*}
    f=10^5\ {\rm Hz},
    \qquad
    \omega=2\pi f.
\end{equation*}
The characteristic electromagnetic velocity used in the
numerical scaling is
\begin{equation}
    U_c
    =
    \frac{
    \mu_m\sigma^2 E_c^2 L_c^3
    }{\mu},
    \label{eq:Uc_appendix}
\end{equation}
with $E_c=3000$ in the numerical implementation.

\bibliography{biblio}

\end{document}